\documentclass[aps,prc,twocolumn,superscriptaddress,nofootinbib,showpacs,showkeys,preprintnumbers]{revtex4-1}

\usepackage[colorlinks=true, pdfstartview=FitV, linkcolor=red, citecolor=blue, urlcolor=blue]{hyperref}
\usepackage{graphicx}
\usepackage{dcolumn}
\usepackage{bm}
\usepackage{xspace}
\usepackage{dirtytalk}
\usepackage{lipsum}
\usepackage{natbib}
\usepackage{lineno}

\usepackage{amsmath}
\usepackage[TS1,T1]{fontenc}
\usepackage{textcomp}
\usepackage[dvipsnames]{xcolor}

\newcommand{\chun}[1]{#1}
\newcommand{\mike}[1]{#1}

\newcommand{\R}[3]{\mbox{$#1{\mathcal{R}}_{#2}^{\hat{#3}}$}\xspace}
\newcommand{\Rfluid}[1]{\mbox{$\R{}{\rm fluid}{#1}$}}
\newcommand{\Rlambda}[1]{\mbox{$\R{}{\Lambda}{#1}$}}
\newcommand{\Rbarlambda}[1]{\mbox{$\R{\overline}{\Lambda}{#1}$}}
\newcommand{\RbarAntilambda}[1]{\mbox{$\R{\overline}{\overline{\Lambda}}{#1}$}}
\newcommand{\RbarNR}[1]{\mbox{$\R{\overline}{\rm NR}{#1}$}}

\begin{document}

\title{Vortex rings from high energy central p+A collisions}

\author{Michael Annan Lisa} \affiliation{The Ohio
  State University, Columbus, Ohio, USA} 
\author{Jo\~{a}o Guilherme Prado Barbon}\affiliation{Instituto de F\'{i}sica Gleb Wataghin, Universidade Estadual de Campinas, Campinas, Brasil}
\author{David Dobrigkeit Chinellato}\affiliation{Instituto de F\'{i}sica Gleb Wataghin, Universidade Estadual de Campinas, Campinas, Brasil}
\author{Willian Matioli Serenone}\affiliation{Instituto de F\'{i}sica Gleb Wataghin, Universidade Estadual de Campinas, Campinas, Brasil}
\author{Chun Shen}
\affiliation{Department of Physics and Astronomy, Wayne State University, Detroit, MI 48201, USA}
\affiliation{RIKEN BNL Research Center, Brookhaven National Laboratory, Upton, NY 11973, USA}
\author{Jun Takahashi}\affiliation{Instituto de F\'{i}sica Gleb Wataghin, Universidade Estadual de Campinas, Campinas, Brasil}
%
\author{Giorgio Torrieri}\affiliation{Instituto de F\'{i}sica Gleb Wataghin, Universidade Estadual de Campinas, Campinas, Brasil}

%

\begin{abstract}
Relativistic p+A collisions may produce droplets of quark gluon plasma (QGP) that quickly develop a toroidal vortex structure similar to that of an expanding smoke ring. We present viscous relativistic hydrodynamic calculations of ultra-central p+A collisions and develop an experimental observable to probe the structure, correlating the polarization and momentum of hyperons emitted from the collision. \mike{This effect is robust against
changes in the definition of vorticity used to calculate
the polarization.}
Experiments at RHIC and LHC may test the existence and strength of the vortex toroids, bringing new evidence to bear on the question of collectivity in the smallest QGP droplets.
\end{abstract}


\maketitle

A cylindrically symmetric volume of fluid in which the longitudinal velocity
  of a cell depends on radius will develop toroidal vorticity structures~~\cite{doi:10.1080/14786446708639824}.
Such vortex rings are ubiquitous in fluid dynamics.
A familiar example is a gentle puff of air from a human mouth.
Due to surface friction with the lips, the air at the outer radius of the cylinder receives less
  of a longitudinal impulse than the air in the middle;
  the resulting vortex rings are clearly visible if smoke is present in the expelled air.
  
\begin{figure}
    \centering
    \includegraphics[width=0.25\textwidth]{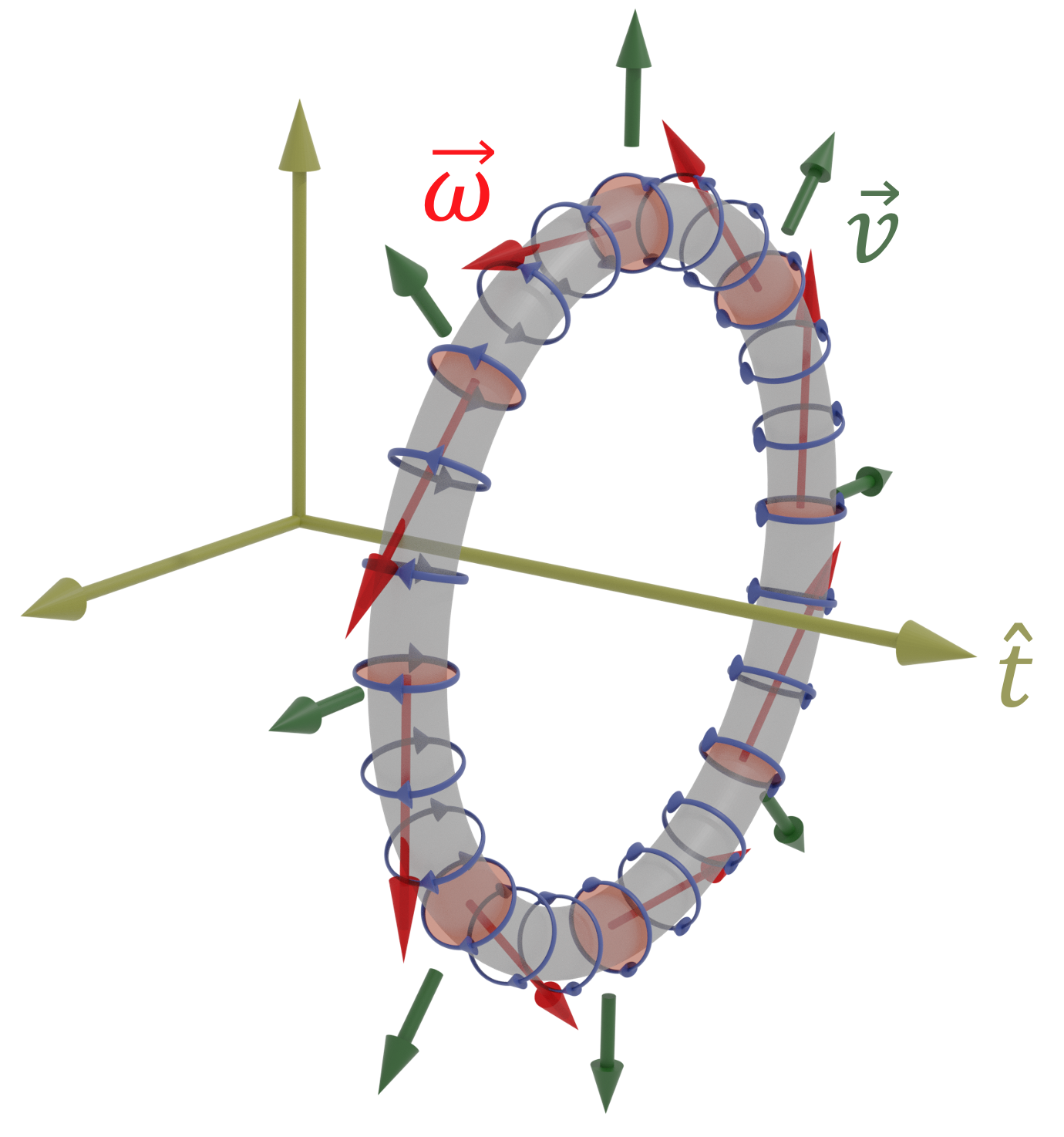}
    \caption{The vortex structure of an expanding smoke ring.}
    \label{fig:SmokeRing}
\end{figure}

The vortical structure of the expanding smoke ring sketched in
  figure~\ref{fig:SmokeRing} may be naturally quantified by
  \begin{linenomath*}
\begin{equation}
\label{eq:RNR}
    \RbarNR{t}
    \equiv
    \left\langle
    \frac{\vec{\omega}_{\rm NR}\cdot\left(\hat{t}\times\vec{v}_{\rm cell}\right)}
    {|\hat{t}\times\vec{v}_{\rm cell}|}
    \right\rangle_{\phi},
\end{equation}
\end{linenomath*}
where $\vec{\omega}_{\rm NR}=\tfrac{1}{2}\vec{\nabla}\times\vec{v}$ is
  the non-relativistic vorticity and
  $\vec{v}_{\rm cell}$ is the velocity of one fluid "cell" of
  smoke.
The axis of the ring is $\hat{t}$ and the structure is averaged
  over azimuthal angle about the ring axis.

In this paper, we investigate the possibility that analogous toroidal vortex structures may appear in 
  a small droplet of quark-gluon plasma created in a
  central ultra-relativistic collision between a proton and a heavy ion.
It is well-established that high-energy collisions between two heavy nuclei have been shown to form a "nearly perfect fluid"~\cite{Heinz:2013th}
  of quarks and gluons.
Recently, the possibility of hydrodynamic collectivity
  in p+A collisions has been the focus of intense
  theoretical and experimental research~\cite{doubleridgeALICE,doubleridgeATLAS,Acharya:2019vdf,Loizides:2016tew, Nagle:2018nvi, Shen:2020mgh}, as these collisions produce anisotropic flow signatures
  similar to those observed from A+A collisions.
However, 
  non-hydrodynamic physics can contribute significantly to such observables in small systems~\cite{Loizides:2016tew,Acharya:2019vdf,Giacalone:2020byk}.

Detailed hydrodynamic calculations accurately reproduce measured anisotropies for heavy systems,
  but the agreement quantitatively deteriorates with decreasing system size~\cite{Schenke:2020mbo}.
It is unclear whether this reflects the need to tune the fine details of the calculation or a fundamental problem with a hydrodynamic
  approach for this system.
By probing non-trivial flow structure involving both transverse and longitudinal directions,
measurements sensitive to vortex-ring structures may provide valuable insight into this issue.
Below, we use state-of-the-art three-dimensional relativistic viscous hydrodynamic calculations~\cite{Schenke:2010nt} to simulate the
  development of the fluid flow field; we suggest an experimental observable
  to probe for vortex rings; and we make predictions for different initial conditions and vorticity definitions.

  \begin{figure}
    \centering
    \includegraphics[width=0.49\textwidth]{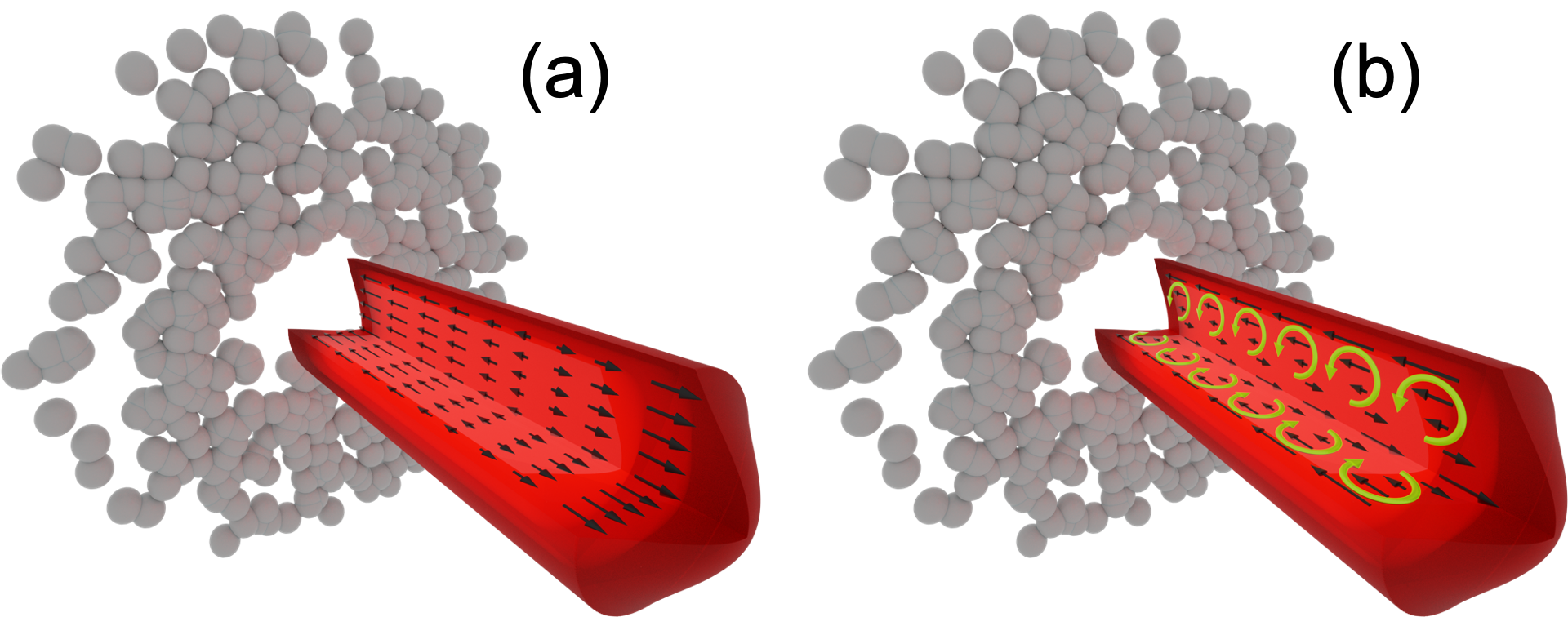}
    \caption{Sketch of the initial conditions for the plasma formed just
    after a proton drills through the center of a gold nucleus.
      The thermalized fluid tube is initialized with two possible flow patterns.
      Left: initial condition (a): a boost-invariant flow distribution
      with more matter in the Au-going direction.
      Right: initial condition (b): the edges of the cylinder flow more
      in the Au-going direction than do fluid cells at the center of the 
      cylinder.
      Transverse flow rapidly develops hydrodynamically, but is not
      present in the initial condition.
      \label{fig:cartoon}
    }
  \end{figure}

To focus our discussion, we ignore the ``lumpy'' structure of the colliding nuclei~\cite{Alver:2010gr} and consider
  completely central collisions between smooth nuclei, as sketched in figure~\ref{fig:cartoon}.
Smooth initial energy density profiles are generated by averaging over many Monte-Carlo collision events with impact parameter $b = 0$.
A boost-invariant initial flow distribution, sketched in panel (a) of the figure, is often assumed for both
  symmetric (e.g. Au+Au) and asymmetric (e.g. p+Au) collisions~\cite{Shen:2020jwv}.
However, if a cylinder of thermalized fluid is quickly produced as the proton drills through the heavy nucleus, the longitudinal
  velocity distribution may feature a radial gradient~\cite{Voloshin:2017kqp} as indicated by panel (b) in the figure.
In both cases, the initial energy density and longitudinal flow velocity 
  are parameterized to ensure local energy and longitudinal momentum conservation at every position in the transverse plane~\cite{Shen:2020jwv}. 
These constraints are essential to ensure the initial orbital angular 
  momentum from the collision geometry is smoothly mapped to fluid dynamic variables.
In scenario (a), the longitudinal flow velocity
  is set equal to the spacetime rapidity of the cell, i.e. the Bjorken flow profile.
The energy density 
  flux tube was shifted forward or backward according to the net longitudinal
  momentum.
In scenario (b),  the net longitudinal momentum is distributed to a non-zero local longitudinal flow $u^\eta$ 
which depends on transverse position but is 
independent of the space-time rapidity $\eta_s$.

\chun{At any given transverse position $(x, y)$, we assume the initial energy-momentum current has the form $T^{\tau \mu} = e(x, y, \eta_s) (\cosh[y_L(x, y)], 0, 0, \sinh[y_L(x, y)]/\tau_0)$, where the longitudinal flow rapidity $y_L(x, y) = f y_\mathrm{CM}(x, y)$ with $y_\mathrm{CM}(x, y) = \mathrm{arctanh} \left[ \frac{T_A - T_B}{T_A + T_B} \tanh(y_\mathrm{beam}) \right]$ encoding the net longitudinal momentum. We use $f = 0$ in scenario (a) and $f = 1$ in scenario (b). For symmetric Au+Au collisions at $b = 0$, $T_A = T_B$ and $y_\mathrm{CM} = 0$, which leads to the same initial flow rapidity for both scenarios. Here $T_{A(B)}(x, y)$ is the nuclear thickness function and $y_\mathrm{beam}$ is the beam rapidity. The local energy density $e(x, y, \eta_s)$ has a flux-tube like profile centered at $y_\mathrm{CM} - y_L$ along the longitudinal direction, $e(x, y, \eta_s) = N \exp \left[- \frac{(\vert \eta_s - (y_\mathrm{CM} - y_L) \vert - \eta_0)^2}{2 \sigma_\eta^2} \right] \theta (\vert \eta_s - (y_\mathrm{CM} - y_L) \vert - \eta_0)$ with $N$ being a normalization factor~\cite{Shen:2020jwv}.}

The geometry of initial condition (b) resembles a recent experiment of 
  Takahashi, {\it et al},~\cite{Takahashi:2016nature}
  in which mercury flowed through a cylindrical tube.
Surface friction with the wall induced an azimuthally oriented vorticity structure 
  with a strength that increased with radius.
Spin-orbit coupling produces an observable electron polarization proportional to the
  local fluid vorticity~\cite[][supp. info.]{Takahashi:2016nature} 
  $\vec{\omega}_{\rm NR}$.

\mike{In an equilibrium ansatz, the fluid}
 vorticity may be probed by measuring the spin polarization of
  $\Lambda$ hyperons through their parity-violating decay mode~\cite{Liang:2004ph,Becattini:2020ngo}.
However, the situation is more complicated in this case, for two
  reasons.
\mike{Firstly, the system requires a fully relativistic treatment
  in which the vorticity is a four-dimensional rank-2 tensor.
The most commonly used is the so-called thermal
  vorticity,
\begin{linenomath*}
\begin{equation}
\label{eq:omegaThermal}
  {\omega}_{\rm th}^{\mu\nu} \equiv \tfrac{1}{2}
  \left[\partial^{\nu}\left(u^{\mu}/T\right) -
  \partial^{\mu}\left(u^{\nu}/T\right) \right] .
\end{equation}
\end{linenomath*}
However, other definitions are possible~\cite{Becattini:2015ska}, including the
  kinetic vorticity,
\begin{linenomath*}
\begin{equation}
\label{eq:omegaKinetic}
  {\omega}_{\rm kin}^{\mu\nu} \equiv \tfrac{1}{2}
  \left[\partial^{\nu}\left(u^{\mu}\right) -
  \partial^{\mu}\left(u^{\nu}\right) \right] ,
\end{equation}
\end{linenomath*}
and the temperature (or "T") vorticity,
\begin{linenomath*}
\begin{equation}
\label{eq:omegaT}
  {\omega}_{\rm T}^{\mu\nu} \equiv \tfrac{1}{2}
  \left[\partial^{\nu}\left(Tu^{\mu}\right) -
  \partial^{\mu}\left(Tu^{\nu}\right) \right] .
\end{equation}
\end{linenomath*}
While it has been argued~\cite{Becattini:2007nd,Becattini:2013fla} that the connection
  between polarization and $\omega_{\rm th}$ is on the firmest theoretical footing,
  using $\omega_{\rm T}$ to calculate longitudinal polarization in Au+Au
  collisions agrees best with experimental observations~\cite{Wu:2019eyi},
  though agreement may be obtained in other ways, as well~\cite{Florkowski:2019voj,Liu:2019krs,Fu:2021pok,Becattini:2021iol}.
In  equations~\ref{eq:Freezeout} and~\ref{eq:Sprop} 
  below, $\omega=\omega_{\rm th}$.
Predictions using the other vorticities are trivially obtained by
  substituting $\omega_{\rm kin}/T$
  or $\omega_{\rm T}/T^2$.
As we discuss below, our predictions for observable polarization in p+A collisions
  are qualitatively similar for all types of vorticity.
}

Secondly, unlike the electrons in the Takahashi experiment, in the
   hydrodynamic paradigm of high energy collisions,
   hadrons are not part of the evolving
  fluid but rather "freeze out" of it, on a hypersurface $\Sigma$ in space
  and time~\cite{Huovinen:2012is, Ahmad:2016ods}
The observable hyperon polarization is dictated by the fluid vorticity
  distribution on this hypersurface according to~\cite{Becattini:2013fla}
    \begin{linenomath*}
\begin{equation}
  \label{eq:Freezeout}
   S^\mu(p)= - \frac{1}{8m} \epsilon^{\mu\rho\sigma\tau} p_\tau 
 \frac{\int d\Sigma_\lambda p^\lambda n_F (1 -n_F) \omega_{\rho\sigma}}
  {\int d\Sigma_\lambda p^\lambda n_F} ,
\end{equation}
\end{linenomath*}
where $n_F$ is the Fermi-Dirac distribution.
Below, we present calculations of $\Lambda$ spin distributions
  using equation~\ref{eq:Freezeout}, with the condition that
  hadrons freeze out occurs when 
  the local energy
  density falls below a critical value of $0.5~{\rm GeV/fm^3}$ \cite{Shen:2020jwv, Oliinychenko:2020znl}.

First, however, it is instructive to observe 
  ring structures in the evolving fluid itself.
It can be shown that the momentum-integrated hyperon spin is
  proportional to
    \begin{linenomath*}
\begin{equation}
\label{eq:Sprop}
    S^{\mu} \propto \epsilon^{\mu\rho\sigma\tau}
    {\int d\Sigma_{\lambda}\omega_{\rho\sigma}\left(Au_{\tau}u^{\lambda}+B\Delta_{\tau}^{\lambda}\right)} ,
\end{equation}
  \end{linenomath*}
where $A$ and $B$ are the thermodynamic integrals
  \begin{linenomath*}
\begin{align}
    \label{equation:AB}
    A &\equiv \int\frac{d^3p}{E}\left(p^{\alpha}u_{\alpha}\right)^2 n_{F}\left(1-n_{F}\right) \\
    B &\equiv \tfrac{1}{3} \int\frac{d^3p}{E}p^{\alpha}p^{\beta}\Delta_{\alpha\beta} n_{F}\left(1-n_{F}\right) , \nonumber
\end{align}
  \end{linenomath*}
\mike{and $\Delta_{\mu\nu}\equiv\left(g_{\mu\nu}-u_{\mu}u_{\nu}\right)$.}
  
Hence, for the purpose of illustration, 
  we can define a relevant proxy for vorticity,
    \begin{linenomath*}
\begin{equation}
    \label{eq:hybridOmega}
    \Omega^{\mu} \equiv -\epsilon^{\mu\rho\sigma\nu}\omega_{\rho\sigma}
    \left(u_{\nu} u^\alpha + C \Delta_{\nu}\,^{\alpha}\right) n_\alpha,
\end{equation}
  \end{linenomath*}
where $n^\alpha \equiv d\Sigma^\alpha/|d\Sigma^\alpha|$ is the normal vector of 
  the fluid cell. To examine ``snapshots'' of the fluid at fixed values of proper
  time $\tau$, the fluid cell normal vector is purely time-like 
  $n^\alpha = (1, 0, 0, 0)$.
Here the coefficient $C=\tfrac{B}{A}$ is a constant with a value between 0 (non-relativistic limit)
and $\tfrac{1}{3}$ (ultrarelativistic limit).
For $\Lambda$ hyperons, $C=0.11$ (0.18) for a temperature of 0.15~GeV (0.30~GeV).

\begin{figure*}
    \centering
    \includegraphics[width=0.75\textwidth]{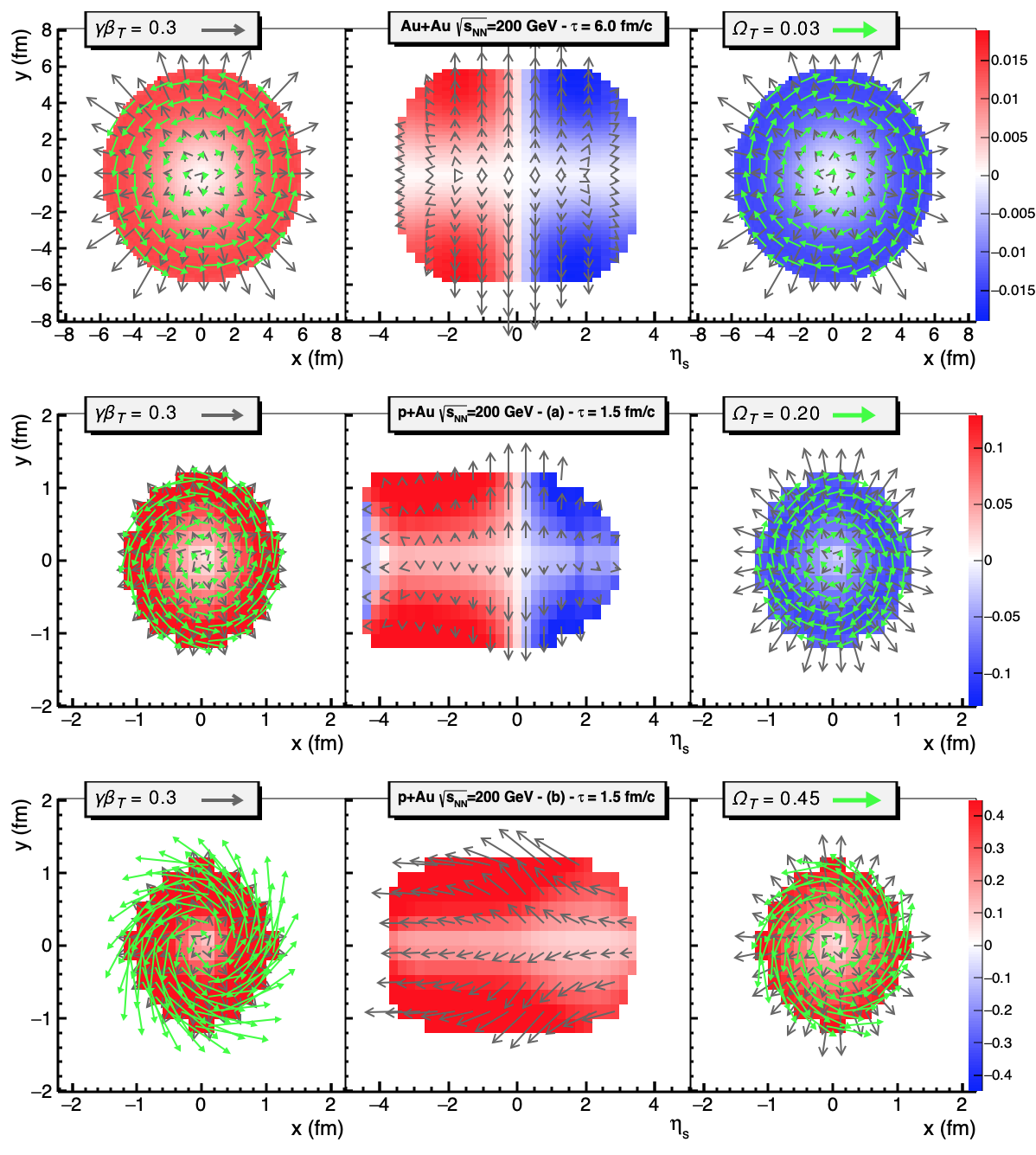}
    \caption{Fluid structure from $\sqrt{s_{\rm NN}}=200$~GeV
    ultracentral collisions between
    smooth Au+Au ions
    (top row) and p+Au collisions with initial conditions (a) (middle)
    (b) (bottom) are shown, roughly halfway through the
    system evolution.
    Black arrows show the average flow velocity, while green
    arrows show the transverse components of the
    vorticity proxy $\Omega$, defined in 
    equation~\ref{eq:hybridOmega} with $C=0.15$.
    The color scale indicates \Rfluid{z},
    defined in equation~\ref{eq:Rfluid}.
    The left (right) panel in each row shows the transverse (to
    the beam direction) plane
    projected over $\eta_s<0$ ($\eta_s>0$).
    For the p+Au collisions,
    the proton travels in the $+\eta_s$ direction.
    The middle panel shows a $x=0$ cross-section of the fluid 
    in the $x-\eta_s$ plane.
    }
    \label{fig:fluidSnapshot}
\end{figure*}

\begin{figure}
    \centering
    \includegraphics[width=0.5\textwidth]{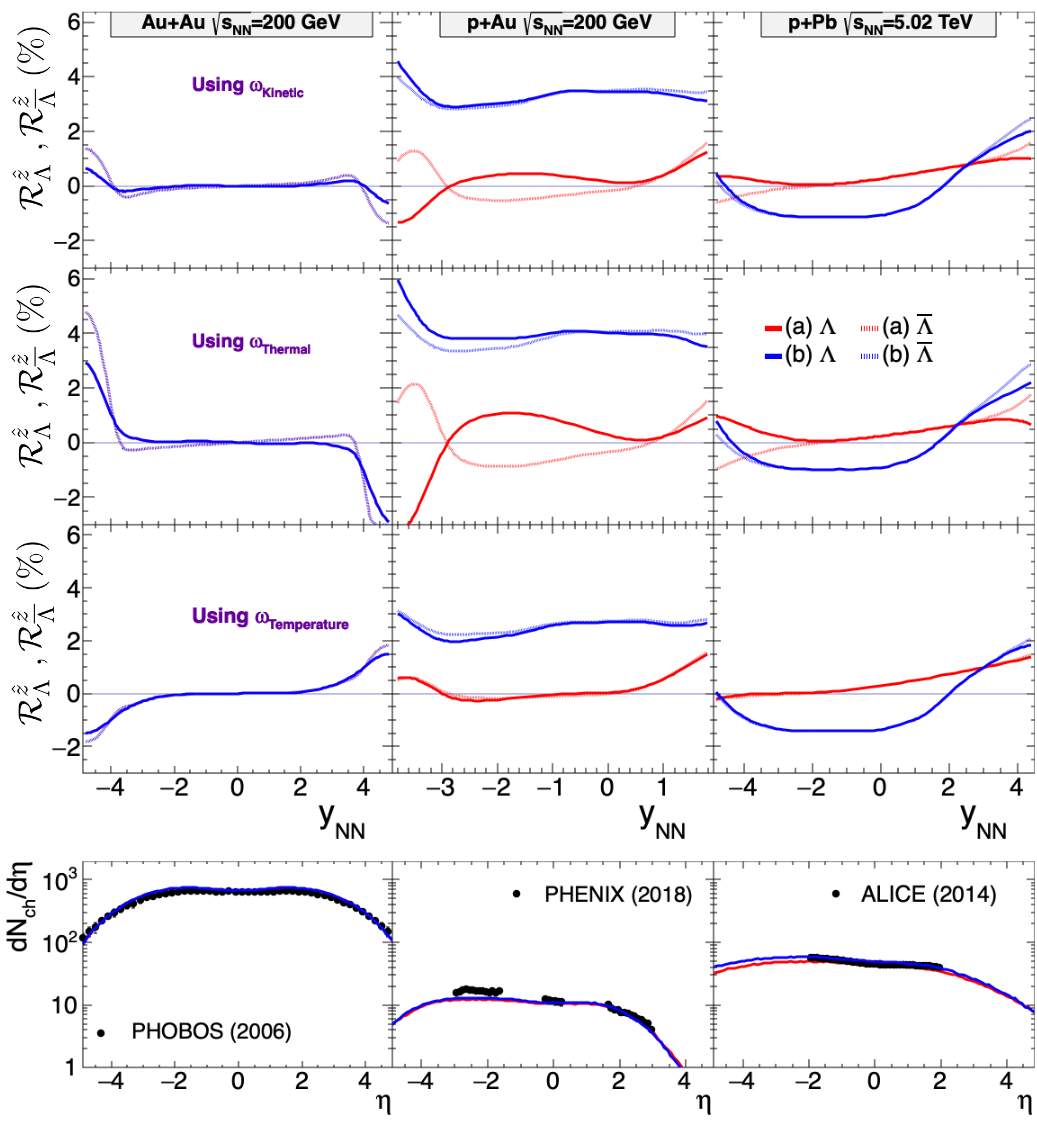}
    \caption{Top panels: \Rbarlambda{z} (solid curve)
    and \RbarAntilambda{z} (dotted curve) are
    plotted as a function of rapidity, integrated over the
    transverse momentum range $0<p_T<3$~GeV/c, for (left)
    ultra-central Au+Au collisions at $\sqrt{s_{\rm NN}}=200$~GeV; (middle)
    p+Au collisions at $\sqrt{s_{\rm NN}}=200$~GeV; and (right) p+Pb collisions at $\sqrt{s_{\rm NN}}=5.02$~TeV.
    Red and blue curves correspond to collisions with initial conditions (a)
    and (b), respectively.
    Particle polarizations calculated from the kinetic, thermal
    and temperature vorticity are shown; see text for details.
    Bottom panels: The pseudorapidity distribution of charged particles.
    Measured distributions~\cite{Back:2005hs,Adare:2018toe,Adam:2014qja} are shown as black symbols.
    All quantities are plotted in the NN frame.
    }
    \label{fig:RLambda}
\end{figure}

Figure~\ref{fig:fluidSnapshot} shows the fluid from an ultracentral
  Au+Au collision and from p+Au collisions with initial conditions
  (a) and (b) shown in figure~\ref{fig:cartoon}.
For the purpose of this illustration, we have set a constant value $C=0.15$, corresponding
  to a temperature of 0.22~GeV.
In the symmetric Au+Au collision, temperature and flow gradients in the transverse
  direction combine with longitudinal flow gradients to produce 
  vortex toroids.
These have been noted previously in  hydrodynamic~\cite{Pang:2016igs,Ivanov:2017dff,Ivanov:2018eej,Ivanov:2019wzg,Ivanov:2020ony,Fu:2020oxj} and 
  transport~\cite{Baznat:2013zx,Teryaev:2015gxa,Baznat:2015eca,Deng:2016gyh,Wei:2018zfb,Xia:2018tes,Zinchenko:2020bbc} simulations for symmetric systems.
\mike{As discussed below, the upper panels of figures~\ref{fig:fluidSnapshot} and~\ref{fig:RLambda}
  show our agreement with these earlier calculations.}

The strength and sense of these vortex toroid structures
  for a given snapshot can be
  naturally quantified analogously to the non-relativistic smoke ring: 
      \begin{linenomath*}
\begin{equation}
\label{eq:Rfluid}
    \Rfluid{t}
    \equiv 
    \frac{\epsilon^{\mu\nu\rho\sigma} \Omega_{\mu} n_{\nu} \hat{t}_{\rho} u_{\sigma}}
    {|\epsilon^{\mu\nu\rho\sigma}n_{\nu} \hat{t}_{\rho} u_{\sigma}|} \quad .
\end{equation}
    \end{linenomath*}
This  reverts to formula~\ref{eq:RNR}
  in the nonrelativistic limit.

In the case under discussion, the ring axis is 
  the proton beam direction, $\hat{t}^\rho = \hat{z}^\rho =(0,0,0,1)$. 
(Rings formed by hard-scattered partons
  losing energy
  in the quark-gluon plasma might be best studied by setting
  $\hat{t}$ along the jet direction.)
The color in figure~\ref{fig:fluidSnapshot} represents
  $\Rfluid{z}$ for each fluid cell.

The bottom two rows of figure~\ref{fig:fluidSnapshot} show p+Au collisions
  roughly half-way through their evolution
  in the nucleon-nucleon center-of-momentum frame (NN), in which
  the colliding proton and Au nuclei have equal and opposite rapidity.
The boost-invariant initial condition (a) (c.f. figure~\ref{fig:cartoon})
  produces a toroidal vorticity structure similar to that of Au+Au
  collisions.
The pattern is not identical due to differences in the transverse density
  distribution of protons and Au nuclei, and because the matter distribution
  is asymmetric in the p+Au
  case; in particular, the density distribution is heavily weighted towards
  the Au-going direction, $\eta_s<0$~\cite{Schenke:2020mbo}.
In both cases, \Rfluid{z} changes sign
  at spacetime rapidity $\eta_s=0$.

Central p+Au collisions initialized with condition (b) show a very 
  different pattern.
A continuous vortex tube structure, imprinted at thermalization to simulate
  the shear at the surface of the fluid, evolves with time, with large
  $\Rfluid{z}>0$ that does not
  change sign, even at $\eta_s=0$.

Experimentally, the spin distribution can be probed with
  the observable 
      \begin{linenomath*}
\begin{equation}
\label{eq:Rlambda}
    \Rlambda{t}
    \equiv 
    \frac{\epsilon^{\mu\nu\rho\sigma}S_{\mu} n_{\nu} \hat{t}_{\rho}p_{\sigma}}
    {|S||\epsilon^{\mu\nu\rho\sigma}n_{\nu}\hat{t}_{\rho}p_{\sigma}|} \quad . 
\end{equation}
    \end{linenomath*}
Again setting $n=(1,0,0,0)$ in the NN frame, the ring structure may be quantified 
    \begin{linenomath*}
\begin{equation}
    \label{eq:RLambda3vectors}
    \Rbarlambda{t} =
    2\left\langle\frac{\vec{S}^{\prime}_{\Lambda}\cdot\left(\hat{t}^{\prime}\times\vec{p}^{~\prime}_{\Lambda}\right)}{|\hat{t}^{\prime}\times\vec{p}^{~\prime}_{\Lambda}|}\right\rangle_{\phi} ,
\end{equation}
    \end{linenomath*}
where  $\hat{t}^{\prime}$ is a purely spacelike trigger direction
and $\phi$ is the azimuthal angle about that direction.
The hyperon polarization is $2\vec{S}^{\prime}_{\Lambda}$, and its momentum is
$\vec{p}^{~\prime}_{\Lambda}$.
The primes ($^\prime$) indicate that the
three-vectors are measured in the NN frame.

Figure~\ref{fig:RLambda} shows the calculated rapidity
dependence, integrated over $p_T$, of \Rbarlambda{z} 
for three systems.
For Au+Au collisions at 200 GeV (top RHIC energy), \Rbarlambda{z} 
is very small near midrapidity and antisymmetric about $y=0$.
For this system, there is no distinction between (a) and (b) by definition.

For p+Au collisions, \Rbarlambda{z} depends
strongly on initial conditions.
The standard, boost-invariant 
Bjorken flow initial scenario (a) results in a small
\Rbarlambda{z}.
However, despite the fact that
  \Rfluid{z} changes sign
  at $\eta_s=0$ (c.f. figure~\ref{fig:fluidSnapshot}),
\Rbarlambda{z} has a small positive offset, as the $\eta_s<0$ (Au-going) region 
  dominates the less-dense $\eta_s>0$ region due to thermal smearing at freezeout.
Meanwhile, initial condition (b) results in a relatively $y$-independent
  value of $\Rbarlambda{z}$
 about an order of magnitude larger.
\mike{Especially at lower collision energy, \RbarAntilambda{z} differs somewhat from
  \Rbarlambda{z} due to finite baryon chemical potential effects
  in our calculations which conserve baryon current.}

Interestingly, for the longer-lived  p+Pb  collision at top LHC
  energy (right column of figure~\ref{fig:RLambda}) the hydrodynamic vorticity
  reverses sign late in the evolution, driven by strong transverse flow.
This more complicated evolution also generates a nontrivial $p_T$ dependence, which we will discuss in a longer study.
It is not surprising that both the initial vortex structure ((a) versus (b))
  and the hydrodynamic evolution  affect the final observable, and it highlights
  the importance of constraining several model parameters (e.g. the evolution time)
  simultaneously~\cite{Everett:2020xug, Nijs:2020roc} through comparison with data. 
  
\mike{The top three rows of figure~\ref{fig:RLambda} show \Rlambda{z} using the
  kinetic, thermal, and temperature vorticity.
While the results differ in the details, clearly the effect at the focus
  of this paper here is robust.
The only significant change is seen in Au+Au collisions at forward
  rapidity, in which the sense of the vortex rings induced by temperature
  gradients has a different sign when using the temperature vorticity,
  reminiscent of the sign difference in longitudinal polarization discussed
  above.
}

The lower panels depict calculated pseudorapidity distributions of charged
particles.
The agreement with measurements~\cite{Back:2005hs,Adare:2018toe,Adam:2014qja}
  is reasonable, and the effect of changing (only) the initial flow configuration is small.

%
The hydrodynamic paradigm is most justifiable at
high density, so predictions in the low-density and steeply-falling tails of 
  $dN_{\rm ch}/d\eta$ may be less reliable.
While the exact ``cutoff value'' is somewhat
arbitrary, we suggest to focus on predictions
  in the kinematic range where $dN_{\rm ch}/d\eta\gtrsim10$, as hadronic observables are dominate by hydrodynamic final state effects \cite{Schenke:2020mbo, Giacalone:2020byk}.
For the p+Au collisions at RHIC, this is $-4\lesssim y_{NN}\lesssim2$.
At both RHIC and LHC, 
tracking detectors that measure $\Lambda$ hyperons 
  cover $y_{NN}\approx0$.

Equation~\ref{eq:RLambda3vectors} quantifies hyperon polarization relative to the
  so-called production plane, a phenomenon first observed more than 40 years ago
  in p+p collisions~\cite{Bunce:1976yb}.
Production plane polarization has since been reported over a wide range of 
  energies~\cite{Hauenstein:2016som} in p+p
  as well as proton collisions with heavy nuclei
  at energies up to~$\sqrt{s_{NN}}=41$~GeV~\cite{Abe:1986ew,Lundberg:1989hw,Abt:2006da,Agakishiev:2014kdy}.
These observations have generally been interpreted
  as elementary processes, the QCD analogy of spin-orbit interactions applied to leading partons, which would explain the absence of the effect on $\overline{\Lambda}$.

In those measurements, (1) \Rbarlambda{z} is independent of $\sqrt{s_{NN}}$ for
  relativistic collisions; (2) it is negative at forward rapidity; and
  (3) its magnitude increases with $y_{NN}$.
These systematics contrast with hydrodynamic predictions shown in 
  figure~\ref{fig:RLambda}.
Furthermore, (4) in the lower energy data, $\RbarAntilambda{z}=0$, while in
  a collective fluid, vorticity polarizes all emitted particles and
  $\RbarAntilambda{z}\approx\Rbarlambda{z}$,
  though baryon number conservation effects lead to detailed differences between
  $\Lambda$ and $\overline{\Lambda}$ at RHIC
  energies.
Therefore, an energy scan of p+A collisions may
  reveal the emergence of hydrodynamic collectivity
  from hadronic background.

In summary, if an ultrarelativistic central p+Au collision forms a tiny droplet 
  of quark-gluon
  plasma, it may generate a toroidal vorticity structure
  generically seen in other fluids initialized under similar geometrical conditions.
We have presented the formalism to quantify the strength of this structure in
  a covariant form (\Rfluid{t}) 
  and performed three-dimensional viscous relativistic
  hydrodynamic calculations to evaluate it.
Initializing the simulation with the commonly-used Bjorken flow condition leads
  to a weak toroidal vorticity structure similar to that seen in ultracentral
  Au+Au collisions; here, vorticity is locally generated dynamically by
  temperature and flow gradients.
In what we consider the more natural initial condition, however, the geometry
  of the collision itself generates a vortical structure, which dominates and
  persists through the evolution to hadronic freezeout.
The two scenarios differ in both the strength and space-time rapidity symmetry 
  of \Rfluid{z}.

We have introduced an experimental observable, \Rbarlambda{z}, that
  probes the toroidal
  vorticity structure, correlating the transverse momentum and polarization
  of hyperons.
Under initial condition (b), our calculations indicate $\Rbarlambda{z}\approx3\%$
  over a broad range of rapidity, 
  roughly an order of magnitude larger than
  \Rbarlambda{z} using condition (a).
This is significantly larger than the global polarization signal measured in semicentral Au+Au
  collisions at $\sqrt{s_{NN}}=200$~GeV at RHIC~\cite{STAR:2017ckg,Adam:2018ivw}; measuring \Rbarlambda{z} 
  should be significantly simpler than measuring
  global polarization, as the latter requires
  correlation with an event plane estimated with
  finite resolution.
\mike{Importantly, this conclusion is robust against changes in the
definition of vorticity used in the calculation of polarization.}
  
More investigation is needed to fully understand 
  the novel
  vortical
  structures potentially generated by the 
  unique geometry
  and dynamics of asymmetric subatomic collisions.
This includes including
  effects of fluctuating initial conditions~\cite{Alver:2010gr},
  \mike{particle structure effects during freezeout,
  and non-vortical contributions to the polarization~\cite{Fu:2021pok,Liu:2021uhn,Becattini:2021iol}}.
Varying the size of the light collision partner (e.g.
  using oxygen rather than proton beam) and the energy
  of the collision could reveal the emergence of
  collectivity in the tiniest droplets of quark gluon plasma.

\begin{acknowledgements}
MAL is supported by the U.S. Department of Energy grant DE-SC0020651,  
acknowledges support of the Fulbright Commission of Brazil,
and appreciates the hospitality of Unicamp.
JPB, DDC, WMS, JT and GT were supported by FAPESP projects 17/05685-2 (all), 19/05700-7 (WMS), 19/16293-3 (JPB) and 17/06508-7 (GT).
CS is supported by the U.S. Department of Energy under grant number DE-SC0013460 and the National Science Foundation under grant number PHY-2012922. This research used resources of the high performance computing services at Wayne State University.
GT acknowledges CNPQ bolsa de
 produtividade 301432/2017-1.
\end{acknowledgements}
 

%

\end{document}